\documentclass[12pt]{article}
\usepackage[USletter]{vmargin}
\usepackage[T1]{fontenc}
\usepackage[]{times}
\usepackage{psfig,fancybox,fancyhdr,epsf}
\usepackage[]{fleqn}
\usepackage[dvips]{graphics}
\usepackage[english]{babel}
\begin{document}

\title{\textbf{Non-equilibrium thermodynamics of gene expression and transcriptional regulation}}
\author{\textbf{Enrique Hernández Lemus} $^{1,2}$ \thanks{ehernandez@inmegen.gob.mx} \\ \small{1 Department of Computational Genomics } \\ \small{National Institute of Genomic Medicine}\\ \small{Periférico Sur No. 4124, Torre Zafiro II, Piso 5}\\ \small{Col. Ex Rancho de Anzaldo, Álvaro Obregón 01900, México, D.F.  México}\\ \small{2 Center for the Sciences of Complexity},\\ \small{Universidad Nacional Autónoma de México}\\ \small{Torre de Ingeniería, Piso 6, Circuito Escolar s/n}\\ \small{Ciudad Universitaria, Coyoacán, 04510, México, D.F., México}}
\date{}
\maketitle

\begin{abstract}
In recent times whole-genome gene expression analysis has turned out to be a highly important tool to study the coordinated function of a very large number of genes within their corresponding cellular environment, especially in relation to phenotypic diversity and disease. A wide variety of methods of quantitative analysis have been developed to cope with high throughput data sets generated by gene expression profiling experiments. Due to the complexity associated with \textit{transcriptomics}, specially in the case of gene regulation phenomena, most of these methods are of a probabilistic or statistical nature. Even if these methods have reached a central status in the development of an integrative, systematic understanding of the associated biological processes, they very rarely constitute a concrete guide to the actual physicochemical mechanisms behind biological function and the role of these methods is more on a hypotheses generating line. An important improvement could be done with the development of a thermodynamic theory for gene expression and transcriptional regulation that will build the foundations for a proper integration of the vast amount of molecular biophysical data and could lead, in the future, to a systemic view of genetic transcription and regulation. .

\end{abstract}

\section{Introduction}

Cellular phenotypes are mainly determined by the expression levels of many genes and their products such as enzymes, proteins and so on. One important tool to track down this cellular phenotypic diversity is gene expression analysis.  One hard-to-grasp issue is that the process of gene expression by itself is a complex one, both from the biochemical and thermodynamical points of view \cite{krug}. The transcription of messenger RNA (mRNA) for a given gene from a DNA template often is regulated by different genes and their products. Being this the case, a variety of physicochemical interactions abounds between genetic transcripts abundance and it is a recognized fact that such complex processes are behind the ultimate mechanisms of cell function. Under this scenario, gene expression values are measured under different conditions, either on a simultaneous [steady-state] or serial [dynamics] fashion, in many cases the measurements are then treated as samples from a joint probability distribution. Genome-wide transcriptional profiling, also called Gene Expression Analysis (GEA) has allowed us to go well beyond studying gene expression at the level of individual components of a given process by providing global information about functional connections between genes, mRNAs and the related regulatory proteins. GEA have greatly increased our understanding of the interplay between different events in gene regulation and have pointed out to previously unappreciated biological functional relations, such as the coupling between nuclear and cytoplasmic transcription and metabolic processes \cite{silver}. GEA also revealed extensive communication within regulatory units, for example in the organization of transcription factors into regulatory motifs.\\

The transcriptional behavior for every gene is simultaneously regulated by both, its related chromatin structure and associated transcription factors. In eukaryotes (organisms with a cellular nucleus), for example, genomic DNA is packaged into nucleosomes that are made of DNA and octamers of a class of proteins called \textit{histones}. Another set of proteins called \textit{chromatin modifiers} are able to move the histones all along the DNA chain to expose specific regions and then, replace histones with \textit{specific histone variants} to convert chromatin from a transcriptionally repressed state into a transcriptionally accessible state, hence enabling gene expression. In the case of transcription factors (TFs), they bind at regulatory regions to either activate or repress the transcription of their \emph{target genes}. TFs do this by (respectively) promoting or inhibiting recruitment of RNA polymerase II. TFs also recruit chromatin-modifying enzymes to make their target DNA more accessible to the \emph{transcriptional machinery} (for a more detailed account see section 3). In the past, the different steps involved in the regulation of gene expression-transcription, mRNA processing, nuclear export, translation and degradation - were usually analyzed \emph{in isolation} by using conventional biochemical techniques. This way of looking at things has given the impression that such processes are independent. Former investigations were focused on the mechanisms underlying individual gene expression or in the best scenario the behavior of a small set of genes, rather than exploring regulatory mechanisms that can influence many genes at one time. Systematic studies of genome-wide binding patterns made evident the existence of a great deal of coordinate regulation among TFs. Factors that combinatorialy regulate (on a concomitant way) a particular gene also often coordinately regulate the expression of other genes, potentially even themselves or each other. Given this fact, they are not independent inputs that merge only at a particular promoter, but rather are coupled. Of course these complex phenomena will ultimately affect a thermodynamical description of transcription regulation because the concentrations (expression levels) and chemical potentials of mRNA transcripts are combinatorialy correlated.\\

Nevertheless, even if we are now provided with experimental techniques to measure the behavior of thousands of mRNA transcripts simultaneously, and a great deal of attention has been put on the computational and statistical analysis of such huge amounts of data; the theoretical approach is still looking at the regulatory interactions at a one by one basis. This approach is of course changing towards a more systematic, network-oriented understanding of gene regulation phenomena. One usual means to understand the nature of such intricate phenomena is by using the so-called Gene Regulatory Networks (GRNs). GRNs are powerful graph-theoretical constructs that describe the integrated status of a cell under a specific condition at a given time \cite{aracne}. The complex description given by GRNs consists, generally, in identifying gene interactions from experimental data through the use of theoretical models and computational analysis. Transcriptional network analyses have showed that, instead of being independent, different levels of gene regulation are strongly coupled. In some cases, have been recognized that the factors involved in a specific stage of mRNA transcription can exhibit coordinated behavior, for example, by finding how groups of transcription factors bind cooperatively at many related promoters.\\

\section{Thermodynamics of hybridization}

Understanding the thermodynamical basis of the hybridization process is an important task related to both, the explicit, intrinsic mechanisms of gene expression and its experimental measurement, especially in the case of high throughput technologies such as the gene chips. One initial approach is to calculate hybridization thermodynamics based on the inference of free energies by means of the energetic cost of base-pair opening in the RNA complex \cite{carlon}. This approach has been also applied to understand the selective hybridization processes related to mRNA silencing (gene switching) by means of small interferring RNA molecules (siRNAs) that are RNA molecules that bind (hybridize) to specific mRNA transcripts thus forbidding their ultimate translation into proteins \cite{lu}. In both scenarios the thermodynamic equilibrium and its properties are important to understand and quantify the degree of hybridization, the specificity of it and the steady-state concentration of mRNA transcripts after either the measurement process or the silencing, respectively. In the present paper, we are more interested in the thermodynamics associated with gene expression quantification and profiling in high throughput experiments, since this is (at least at the moment) the ultimate and more accurate laboratory tool to study the mechanism of genetic transcription.\\

According with the Langmuir adsorption model of oligonucleotide hybridization, the specific-hybridization intensity     (or gene-expression signal) for a gene probe as measured by (for example) an Affymetrix-type gene chip \cite{affy} is given by \cite{carlon}:

\begin{equation}\label{specint} \varphi (c, \Delta G) = \frac{A\; c \; e^{-\beta \Delta G}}{1 + c \; e^{-\beta \Delta G}}\end{equation}

where $\beta = \frac{1}{RT}$, $T$ is the local temperature, $R$ is the gas constant, $c$ is the mRNA concentration for this species, $\Delta G$ is the free energy of hybridization, and $A$ is a parameter that sets the scale of the intensity corresponding to the saturation limit $c\gg e^{\beta \Delta G}$. A natural generalization of Eq.  \ref{specint} for a probe $i$ within a set of $M$ gene-probes $(i=1, \dots, M)$ is:

\begin{equation}\label{specint2} \varphi_i (c_i, \Delta G_i) = \frac{A_i\; c_i \; e^{-\beta \Delta G_i}}{1 + c_i \; e^{-\beta \Delta G_i}}\end{equation}

The local chemical potential $\mu_i$ of species $i$ due to the hybridization process is defined as customarily by $\mu_i =\left(\frac{\partial \Delta G_i}{\partial c_i}\right)_{T,P, c_j}$. From Eq. \ref{specint2} it is possible to calculate $\mu_i$ by means of the chain-rule as follows:

\begin{equation}\label{chain} \mu_i = \left(\frac{\partial \Delta G_i}{\partial c_i}\right)_{T,P, c_j} =\left(\frac{\partial \Delta G_i}{\partial \varphi_i}\right)_{T,P, c_j} \left(\frac{\partial \varphi_i}{\partial c_i}\right)_{T,P, c_j}\end{equation}

or in terms of the direct derivatives:

\begin{equation}\label{chain2} \mu_i = \frac{\left(\frac{\partial \varphi_i}{\partial c_i}\right)_{T,P, c_j}} {\left(\frac{\partial \varphi_i}{\partial \Delta G_i}\right)_{T,P, c_j}}\end{equation}

The first derivative is calculated as:

\begin{equation}\label{der1b} \left(\frac{\partial \varphi_i}{\partial c_i}\right)_{T,P, c_j} = A_i \, c_i \, e^{-\beta \Delta G_i} \, \times \, \frac{- \, e^{-\beta \Delta G_i}}{(1 + c_i \, e^{-\beta \Delta G_i})^{2}} + \frac{1}{(1 + c_i \, e^{-\beta \Delta G_i})} \, \times A_i \, e^{-\beta \Delta G_i}\end{equation}

If we re-arrange terms:

\begin{equation}\label{der1c} \left(\frac{\partial \varphi_i}{\partial c_i}\right)_{T,P, c_j} = \frac{A_i \, e^{-\beta \Delta G_i}}{(1+c_i \,e^{-\beta \Delta G_i})} \times \left[ 1 - \frac{c_i \; e^{-\beta \Delta G_i}}{1 + c_i \; e^{-\beta \Delta G_i}} \right]\end{equation}

which can be then expressed in terms of $\varphi_i$ to read:

\begin{equation}\label{der1d} \left(\frac{\partial \varphi_i}{\partial c_i}\right)_{T,P, c_j} = \frac{\varphi_i}{c_i} \left[ 1 - \frac{\varphi_i}{A_i}\right]\end{equation}

Now, in the case of the second derivative in Eq. \ref{chain2}, it is given by:

\begin{equation}\label{der2b} \left(\frac{\partial \varphi_i}{\partial \Delta G_i}\right)_{T,P, c_j} = A_i \, c_i \, e^{-\beta \Delta G_i} \times \left[ \frac{- \, c_i \, e^{-\beta \Delta G_i} (-\beta)}{(1 + c_i e^{-\beta \Delta G_i})^{2}} \right] + \frac{1}{(1 + c_i \,e^{-\beta \Delta G_i})} \times \left[ A_i \, c_i \, e^{-\beta \Delta G_i} (-\beta)\right]\end{equation}

and then simplifies to:

\begin{equation}\label{der2c} \left(\frac{\partial \varphi_i}{\partial \Delta G_i}\right)_{T,P, c_j} = \frac{A_i\, c_i^{2} \, \beta \, e^{-2\beta \Delta G_i}}{(1+ c_i \,e^{-\beta \Delta G_i})^{2}} - \frac{A_i \, c_i \, \beta \, e^{-\beta \Delta G_i}}{1+ c_i \,e^{-\beta \Delta G_i}}\end{equation}

Eq. \ref{der2c}, could also be written in terms of $\varphi_i$

\begin{equation}\label{der2d} \left(\frac{\partial \varphi_i}{\partial \Delta G_i}\right)_{T,P, c_j} = \frac{\beta \, \varphi_i^{2}}{A_i} - \beta \varphi_i \end{equation}

which gives as a result that:

\begin{equation}\label{der2e} \left(\frac{\partial \varphi_i}{\partial \Delta G_i}\right)_{T,P, c_j} = \beta \varphi_i \left[ \frac{\varphi_i}{A_i} -1\right] \end{equation}

recalling Eq. \ref{chain2}, \ref{der1d} and \ref{der2e}, we finally get:

\begin{equation}\label{muif} \mu_i = \frac{\frac{\varphi_i}
{c_i} \left( 1 - \frac{\varphi_i}{A_i}\right)}{\beta \varphi_i \left( \frac{\varphi_i}{A_i} -1 \right)} = \frac{-1}{\beta \, c_i}\end{equation}

or

\begin{equation} \label{muifinal} \mu_i = \frac{-RT}{c_i} \end{equation}

It is interesting to notice that this level of description (two state Langmuir adsorption model) gives an expression (Eq. \ref{muifinal}) for the chemical potential that is equivalent to that of an \emph{ideal gas}, i.e. non-interacting species, for if we calculate the equilibrium \textit{chemical-work} contribution to the free energy, $\Xi_i$, we obtain:

\begin{equation}\label{deltamu} \Xi_i = \int \mu_i \, dc_i = \int \frac{-RT}{c_i} \, dc_i \end{equation}

or

\begin{equation}\label{deltamu2} \Xi_i = {-RT} \; ln \left( \frac{c_i}{c_i^{o}} \right) \end{equation}

This approximation is valid as long as the rate of \textit{cross-hybridized targets} stays low, since if there is only (or mostly) \textit{transcript-specific hybridization}, the chemical species (in this case the different mRNA molecules) could be considered non-interacting. This is a realistic assumption given the low concentrations of every transcript in solution and also the fact that current technologies are very efficient in reducing the rate of unspecific hybridization \cite{affy}.

\section{Transcriptional regulation}

The phenomenon of gene expression (also known as mRNA transcription or simply transcription) is a complex one. There is a set of control mechanisms collectively called \emph{transcriptional regulation} that take the duty to control \emph{when} transcription occurs and also \emph{how much} mRNA is created. The transcription of a given gene by means of the RNA polymerase enzyme (RNApol) can be regulated or controlled by at least five different biochemical mechanisms.

\begin{itemize}

\item There exists a set of proteins called \emph{specificity factors} that alter the specific binding of RNApol to some given \emph{promoter} or set of \emph{promoters}. A promoter is a DNA region located next -technically in the \textsl{upstream cys location} or towards the 5' region of the sense strand- to a gene that facilitates its transcription by making that region easy to recognized by the transcriptional machinery.

\item \emph{Repressors} are DNA-binding proteins whose function is the regulation of the expression of one or more genes by decreasing the rate of transcription. The actual mechanisms involves their attachment to an \emph{operator} hence forbidding the transcription of the adjacent segment of DNA by blocking the pass of RNApol.

\item \emph{Transcription factors} are proteins that bind to specific DNA sequences in order to control the rate of transcription. Transcription factors are able to perform their function alone, or by forming a complex with other proteins. Transcription factors bind to either enhancer or promoter regions of DNA adjacent to the genes that they regulate. Depending on the transcription factor, the transcription of the adjacent gene is either up- (i.e. higher concentrations of the corresponding mRNA) or down-regulated (lower concentrations of mRNA). Transcription factors use a variety of mechanism for the regulation of gene expression. These mechanisms include: stabilize or block the binding of RNApol to DNA, catalyze the acylation or deacylation of DNA. The transcription factor can either do this alone or by recruiting other proteins that possess catalytic activity.

\item The DNA-binding proteins that enhance the interaction of RNApol to a particular promoter region, thus enlarging the expression levels of the associated gene are called \emph{Activators}. Activators perform their work by means of either electrostatic interactions with some sub-units of RNApol (attracting the molecule towards them and hence towards the DNA region they are bound to) or by inducing conformational changes in the structure of DNA that make easier its binding to RNApol.

\item Finally, \emph{Enhancers} are regions in DNA that are able to bound with activators hence bringing promoters to the initiation complex.

\end{itemize}

\section{Non-equilibrium thermodynamics for small reactive systems: the Transcriptional Regulation scenario}

As is already evident from the previous section, the process of gene regulation within a cell is highly complex from the bio-physicochemical point of view. Another source of complexity in the non-equilibrium thermodynamical characterization of such system lies in the fact that a cell is a \emph{small system}, in the sense that its dimensions do not permit an obvious application of the thermodynamic limit. Specifically, the role of fluctuations and stochasticity within such scenarios is not clear. Small systems thermodynamics for equilibrium systems has been studied in the past \cite{hill1, hill2} and some results were even expected to extend to local equilibrium settings within cellular sized biosystems \cite{hillbio}. One important limitation for the development of such theoretical frameworks at that time was the lack of proper experimental settings to test their hypotheses. Nevertheless, with the development of modern techniques, such as microscopic manipulation by means of atomic force microscopes, optical tweezers and cold-traps this situation has become less of a limitation. In the meantime, theories have been developed to explain several results. These include mesoscopic thermodynamical approaches \cite{rubi, ritort} and also studies made by means of the so-called fluctuation theorems \cite{evans, galco, jarzy, crooks}. Some of these theoretical results have been even experimentally tested.

\subsection{Fluctuation phenomena in non-equilibrium systems}

To have a better idea of the role of large local fluctuations in small systems, let us recall an ideal gas composed by N particles. The total energy of the system is a Gaussian distributed random variable with average $<\epsilon> \sim N k_B T$ and variance $\sigma^{2}_{\epsilon} \sim N k_B^{2}T^{2}$. In that (general) case the fluctuations of the system are proportional to $N^{-\frac{1}{2}}$. This means that for systems of size $N \approx {\cal O} [1]$ they are comparable (and thus important !), whereas for a system with $N \approx {\cal O} [10^{23}]$ these same fluctuations become negligible. An interesting case of study is the cell behavior of the RNApol molecule already mentioned. As we have said RNApol is an enzyme that moves along the DNA to produce a newly synthesized mRNA molecule. It has been mentioned that RNApol extracts energy from its surrounding thermal bath (i.e. the cellular environment) to move, and at the same time uses bond hydrolysis to insure that only thermal fluctuations that lead to \emph{forward} movement are captured. RNApol then serves as an out-of-equilibrium thermal rectifier. The complex dynamics behind even this (relatively) simple model of transcription demonstrate the necessity for a non-equilibrium thermodynamical characterization that includes the possibility to deal with fluctuations in small systems.\\

A very important concept in the non-equilibrium fluctuations setting is that of a \emph{control parameter}. Roughly speaking, a control parameter is a variable that must be specified to define on an unambiguous manner the state of a non-equilibrium system, i.e. control parameters are non-fluctuating variables. If we call $x_n$ ($n=1  \dots p$) the set of parameters of a non-equilibrium system and  $x_{\gamma}$ is the control parameter. If we vary $x_{\gamma}$, then the total energy of the system will vary accordingly as:

\begin{equation}\label{gamma} dU = \sum_{n \neq \gamma} \left( \frac{\partial U}{\partial x_n} \right)_{x_{\gamma}} dx_n + \left( \frac{\partial U}{\partial x_{\gamma}}  \right)_{x_n} dx_{\gamma}\end{equation}

One can see that the first term(s) correspond to the variation of energy as a result of internal configurations (we naïvely call this the \emph{heat}) whereas the second term is the energy change due to an external perturbation (that is the \emph{work}). Of course this formulation implies the experimental difficulty of finding an appropriate (natural) control parameter without disturbing (too much) the system. However, since there is a presence of thermal rectification phenomena in non-equilibrium small systems, Eq. \ref{gamma} will serve as a basis for the extended irreversible thermodynamical description below.

\subsection{Mesoscopic non-equilibrium thermodynamics}

As it has already been said, systems outside the realms of the thermodynamic limit are characterized by large fluctuations and hence stochastic effects. The classic thermodynamic theory of irreversible process (CIT) \cite{degroot} gives a rough, \emph{coarse grained} description of the systems, one that ignores all the details of the molecular nature of matter, hence studying it as a continuum media by means of a phenomenological field theory. As such CIT is not suitable for the description of small systems because fluctuations, ignored by CIT could become the dominant factor in the system's dynamical evolution and response. Nevertheless, in many instances (such as the present case of gene expression regulation) it would be desirable to have a thermodynamic theoretical framework to study such so-called \emph{nano-systems}. One possible way to do so is by considering the stochastic nature of the time evolution of small non-equilibrium systems. This is the approach followed by Mesoscopic Non-Equilibrium Thermodynamics (MNET) \cite{rubi}. MNET for small systems could be understood as an extension of the equilibrium thermodynamics of small systems developed by Hill and co-workers \cite{hill1, hill2, hillbio}. \\

The way in which stochasticity is coming into play is by means of recognizing that scaling down the description of a physical system brings up energy contributions that are usually neglected in thermodynamical descriptions either in equilibrium or outside of it. These contributions take the form of, for example, surface energies and bring in turn a disruption of the canonical view of extensivity. An example used by Hill \cite{hill2} is that of a small cluster of N identical particles for which the equilibrium Gibbs energy is given as: $G = \mu N + a N^{\beta}$ with $\mu$ the chemical potential, $a$ an arbitrary adjustig function and $\beta <1$ a \emph{size-effect} exponent. Here, the second term represents these energies that are usually disregarded whose effects become negligible for very large $N$ since the first term becomes dominant. In this way at the thermodynamic limit one gets the usual $G = \mu N$ relation. It is then possible to treat the Gibbs energy as a fluctuating quantity. Of course we can \textit{adjust} the definition of the chemical potential to account for these effects.\\

Defining

\begin{equation}\label{murara} \widehat{\mu} = \mu +a N^{\beta -1}\end{equation}

It is possible to recover the standard Euler relation $G = \widehat{\mu} N$. However, one must be cautious since even if $\widehat{\mu}$ accounts for the actual energy potential involved in the thermochemical description of such a small system. It is NOT a canonical chemical potential, since for instance, it does not a give rise to an extensive thermodynamical description. Of course in the thermodynamic limit $\widehat{\mu} \to \mu$.\\

In the same order of ideas, MNET was developed to characterize non-equilibrium small systems. Let us recall that any reduction of the spatio-temporal scale description of a system would entail an increase in the number of non-coarse grained degrees of freedom (we are \textit{looking} at things with more detail as to say). These degrees of freedom could be related with the extended variables in Extended Irreversible Thermodynamics \cite{jourodgar}, but they could also be more microscopic in nature, such as colloidal-particle velocities, orientational states on a quasi-crystal, and so on. Hence, in order to characterize such variables, let us say that there exist a set $\Upsilon = \{\upsilon _i\}$ of such non-equilibrated degrees of freedom. $P(\Upsilon, t)$ is the probability that the system is at a state given by $\Upsilon$ at time $t$. If one assumes \cite{rubi2}, that the evolution of the degrees of freedom could be described as a diffusion process in  $\Upsilon$-space, then the corresponding Gibbs equation could be written as:

\begin{equation}\label{gibes} \delta S = - \frac{1}{T} \int \mu(\Upsilon) P(\Upsilon, t) d \Upsilon \end{equation}

$\mu(\Upsilon)$ is a generalized chemical potential related to the probability density, whose time-dependent expression could be explicitly be written as:

\begin{equation}\label{gibes2} \mu(\Upsilon, t) = k_B T \, ln \frac{P(\Upsilon, t)}{P(\Upsilon)_{equil}} + \mu_{equil} \end{equation}

or in terms of a \textit{nonequilibrium work term} $\Delta W$:

\begin{equation}\label{gibes2} \mu(\Upsilon, t) = k_B T \, ln P(\Upsilon, t) + \Delta W \end{equation}

The time-evolution of the system could be described as a generalized diffusion process over a potential landscape in the space of mesoscopic variables $\Upsilon$. This process is driven by a generalized mesoscopic-thermodynamic force $\frac{\partial}{d\Upsilon}(\frac{\mu}{T})$ whose explicit stochastic origin could be tracked back by means of a Fokker-Planck-like analysis \cite{rubi, rubi2}. MNET seems to be a good candidate theory for describing non-equilibrium thermodynamics for small systems. In fact the aforementioned arguments point out to MNET being a good choice, \emph{provided one has a suitable model} or microscopic means to infer the probability distribution $P(\Upsilon, t)$.\\

One important setting where MNET seems appropriate is the case of activated processes, like a system crossing a potential barrier. Chemical reactions (and biochemical reactions like the ones involved in gene regulation too!) are clearly in this case. According to \cite{rubi2} the diffusion current in this $\Upsilon$-space could be written in terms of a local fugacity defined as:

\begin{equation}\label{fugaz} z (\Upsilon) = \exp \frac{\mu(\Upsilon)}{k_B T} \end{equation}

and the expression for the associated flux it will be:

\begin{equation}\label{fluxz} J = - k_B \, L \, \frac{1}{z} \frac{\partial z}{\partial \Upsilon} \end{equation}

L is an Onsager-like coefficient. After defining a \emph{diffusion coefficient} $D$ and the associated affinity $A= \mu_2 - \mu_1$, the integrated rate is given as:

\begin{equation}\label{flux2z} \overline J = J_o \left( 1 - \exp \frac{A}{k_B T}\right) \end{equation}

with $J_o = D \exp \frac{\mu_1}{k_B T}$.

One is then able to see that MNET gives rise to nonlinear kinetic laws like Eq. \ref{flux2z}. In this context MNET has been applied successfully in the past in biomolecular processes at (or under) the cellular level or description \cite{rubi3}. In that scenario, non-linear kinetics are used to express, for example RNA unfolding rates as \emph{diffusion currents}, modeled via transition state theory, giving rise to Arrhenius-type non-linear equations. In that case the current was proportional to the chemical potential difference (cf. equation 17 of reference \cite{rubi3}), so the entropy production was quadratic in that chemical potential gradient. We will re-examine these kind of dependency later when discussing gene expression kinetics.\\

Since whole-genome transcriptional regulation consists on a (huge) series of biochemical reactions, and many of these has unexplored chemical kinetics, a detailed MNET analysis such as the one described above is unattainable at the present moment. On what follows, we will explore a phenomenologically based approach that nevertheless takes into account (although in a more intuitive, less explicit way) similar considerations as the MNET framework already sketched. This phenomenological approach is based on the Extended Irreversible Thermodynamics assumption of enlargement of the thermodynamical variables space \cite{eit, eit2}.

\subsection{Extended Irreversible Thermodynamics}
We shall start our discussion by assuming that a generalized entropy-like function $\Psi$ exists, which may be written in the form \cite{jourodgar, eumat}:

\begin{equation} {\frac{d \Psi}{dt}} = T^{-1}[{\frac{dU}{dt}} +  p{\frac{dv}{dt}} - \sum_i \mu_i {\frac{dC_i}{dt}} -  \sum_j {\cal X}_j\odot{\frac{d\Phi_j}{dt}}]\end{equation}

 or as a differential form:

\begin{equation} \label{entrocal} d_t \Psi = T^{-1}[d_tU + p d_tv - \sum_i \mu_i d_tC_i - \sum_j {\cal X}_j\odot d_t\Phi_j]\end{equation}

We see that Eq. \ref{entrocal} is nothing but the formal extension of the celebrated Gibbs equation of equilibrium thermodynamics for the case of a multi-component out of equilibrium system. The quantities appearing therein are the usual ones: $T$ is the local temperature, $p$ and $V$ the pressure and volume, etc. $X_{j}$ and $\Phi_{j}$ are extended thermodynamical fluxes and forces. These extended forces and fluxes are the new elements of EIT, the ones that take into account the aforementioned non-local effects.\\

In the case of a multicomponent mRNA mixture at fixed volume and pressure, we will take our set of relevant variables to consist in the temperature $T(\vec r, t)$ and concentration of each gene species $C_i(\vec r, t)$ as the slow varying (classical) parameters set $\cal{S}$ and the \emph{mass flux} of these species $\vec J_i (\vec r, t)$ as fast variables on the extended set $\cal{F}$ so that ${\cal{G}} = \cal{S} \, \bigcup \, \cal{F}$. These latter variables will take into account the presence of inhomogeneous regions (concentration domains formed because of the gene regulatory interactions) and so will correct the predictions based on the local equilibrium hypothesis. The non-equilibrium Gibbs free energy for a mixture of $i= 1 \dots M$, mRNA transcripts at constant pressure, then reads:

\begin{equation} \label{negibb} d_tG = -\Psi d_tT + \sum_i {\mu_i d_tC_i} + \sum_j{{\cal X}_j \odot d_t\Phi_j} \end{equation}

If one is to consider gene expression/regulation as a chemical process, it must be useful to write things up in terms of the extent of reaction $\xi$, hence $(d_t G)_{T,P, \Phi_j} = \sum_i \mu_i^{\dagger} d_t N_i$ is rewritten by means of the definition of the so-called \textsl{stoichiometric coefficient} $\nu_i = \frac{\partial N_i}{\partial \xi}$ and of the \textsl{chemical affinity} ${\cal A} = \sum_i \mu_i^{\dagger} \nu_i$. The stoichiometric coefficients and the chemical affinities could be defined likewise for a set of ($k=1\dots R$) regulatory interactions (considered as \textit{chemical reactions}) as follows:

\begin{equation}\label{dege} d_tG = -\Psi d_tT + \sum_k {\cal A}_k d_t \xi_k + \sum_j{{\cal X}_j \odot d_t\Phi_j} \end{equation}

or

\begin{equation}\label{dege2} d_tG = -\Psi d_tT + \sum_k \left[\sum_i \mu_{i,k}^{\dagger} \nu_{i,k} \right] d_t \xi_k + \sum_j{{\cal X}_j \odot d_t\Phi_j} \end{equation}

\subsection{Mean field approach}

In many cases the explicit stoichiometry of the regulatory interactions is unknown and in the vast majority of the already studied cases the reactions are given on a one-to-one basis, i.e. one molecule of a transcription factor on each gene-transcription site (or one molecule of each kind of transcription factor in the case of multi-regulated gene targets). Given this, at the moment we will assume $\nu_i = 1; \, \forall i$. In this \textit{diluted} case we have that the extent of each reaction is then proportional to the concentration rate of change and we recover the non-reactive regime similar to that given by Eq. \ref{negibb}. It is important to stress that this approximation is not a disparate one given the fact that the usual DNA/RNA concentrations within the cells are in the picomolar-nanomolar regime. Also, of the almost 30,000 different genes in humans just a small number of these (about 1000-1500) are known to be transcription factors. Nevertheless in order to take into account the scarce yet important gene regulatory interactions (albeit in an indirect manner) we retain the generalized force-flux terms to get:

\begin{equation}\label{dege3} d_tG = -\Psi d_tT + \sum_i {\mu_i d_tC_i} + \sum_j{{\cal X}_j \odot d_t\Phi_j} \end{equation}

Since gene regulation occurs within the cell, it is possible to relate an internal \emph{work} term with the regulation process itself, being this a \emph{far from equilibrium} contribution. This non-local contribution is given by the generalized force-flux term (third term in the r.h.s. of Eq. \ref{dege3}). This is so as gene regulation often does not occur \emph{in situ} and also since is the only way to take into account (albeit indirectly) the changes in the local chemical potentials that cause the long tails in the fluctuations distributions characteristic of non-equilibrium small systems (e.g. cells). The term relating mRNA \emph{flows} due to transcriptional regulation could be written as a product of extended fluxes $\Phi_j$ and forces ${\cal X}_j$. Here $j=1, \dots M$ refers to the different mRNA species being regulated, that is, indexes $i$ and $j$ refer to the very same set of mRNA transcripts but in one case ($i$) we take into account their local equilibrium behavior (as given by their independent chemical potentials and average local concentrations) and in the other case ($j$) we are interested in their highly fluctuating (far from equilibrium) behavior as given by the term $\sum_j{{\cal X}_j \odot d_t\Phi_j}$\\

Now we are faced with the task to propose a form for the extended fluxes and forces within this highly fluctuating regime, that at the same time allow for experimental verification, is simple enough to be solved and it is compatible with the axioms of extended irreversible thermodynamics. As a first approach, we are proposing a system of linear (in the forces) coupled fluxes with memory that was used to successfully characterize another highly fluctuating system, a fluid mixture near the critical point \cite{ehl1}.\\

The constitutive equations are,

\begin{equation} \label{phi1} \vec \Phi_j (\vec r,t) =  \sum_k \int_{-\infty} ^{t} \lambda^{\Phi}_{j,k} \, \vec u \, e ^{\frac {(t'-t)}{\tau^{\Phi}_j}} \mu_{j,k}(\vec r,t') dt'\end{equation}

\begin{equation}\label{equis1} \vec X_j (\vec r,t) = \int_{-\infty} ^{t} \lambda^{X}_j \, e ^{{(t''-t)}\over{\tau^{X}_j}} \vec \Phi_j (\vec r,t'') dt'' \end{equation}

The $\lambda$'s are time-independent, but possibly anisotropic amplitudes, $\vec u$ is a unit vector in the direction of mass flow (the nature of $\vec u$ will not affect the rest of our description, since we will be dealing with the magnitude of the mass flux $|\vec \Phi_j|$) and $\tau$'s are the associated relaxation times considered path-independent scalars. Since we have a linear relation between thermodynamic fluxes and forces some features of the Onsager-Casimir formalism will still hold. This will be especially important when considering cross-regulatory interactions. An interesting question for future research will be whether gene transcription interactions as modeled here obey Onsager's reciprocal relations. \\

Dynamic coupling is given by Eq. \ref{phi1} and \ref{equis1}, nevertheless due to the fact that actual transcription measurement experiments are made either on homeostasis (steady state) settings or within time series designs with intervals several orders of magnitude larger than the associated relaxation times (which are of the order of a few molecular collision times) it is possible to take the limits $\tau^{\Phi}_j \to 0$ and $\tau^{X}_j \to 0$, then the integrals become evaluated delta functions to give:

\begin{equation} \label{phi2} \vec \Phi_j (\vec r,t) =  \vec u \sum_k \lambda^{\Phi}_{j,k} \, \mu_{j,k}(\vec r,t) \end{equation}

\begin{equation}\label{equis2} \vec X_j (\vec r,t) = \lambda^{X}_j \,  \vec \Phi_j (\vec r,t) \end{equation}

It is important to notice that in the future, it will must surely became possible to experimentally measure gene expression in time intervals much shorter (maybe even in real time). In that case, the appropriate theoretical setting will be given by Eq. \ref{phi1} and \ref{equis1} that represent the dynamic nature of the coupling better than Eq. \ref{phi2} and \ref{equis2}.\\

Also due to the spatial nature of the experimental measurements (either RNA blots or DNA/RNA chips measure space-averaged mRNA concentrations) it is possible to work with the related scalar quantities instead, to give:

\begin{equation} \label{phi3} \Phi_j (\vec r,t) =  \sum_k \lambda^{\Phi}_{j,k} \; \mu_{j,k}(\vec r,t) \end{equation}

\begin{equation}\label{equis3} X_j (\vec r,t) =  \lambda^{X}_j \,  \Phi_j (\vec r,t) \end{equation}

Substituting Eq. \ref{phi3} and \ref{equis3} into Eq. \ref{dege3} one gets:

\begin{equation}\label{dege4} d_tG = -\Psi d_tT + \sum_i {\mu_i d_tC_i} + \sum_j \sum_k \left(\lambda^{\Phi}_{j,k} \, \mu_{j,k}\right) d_t \left( \lambda^{X}_j \,  \Phi_j \right) \end{equation}

Assuming the generalized transport coefficient $\lambda^{X}_j$ to be independent of the flux $\Phi_j$ we are able to write:

\begin{equation}\label{dege5} d_tG = -\Psi d_tT + \sum_i {\mu_i d_tC_i} + \sum_j \sum_k \left(\lambda^{\Phi}_{j,k} \, \mu_{j,k}\right) \lambda^{X}_j d_t \Phi_j \end{equation}

Or in terms of the transcription regulation \emph{chemical potentials} $\mu_{j,k}$

\begin{equation}\label{dege6} d_tG = -\Psi d_tT + \sum_i {\mu_i d_tC_i} + \sum_j \sum_k \left(\lambda^{\Phi}_{j,k} \, \mu_{j,k}\right) \lambda^{X}_j \left( \lambda^{\Phi}_{j,k} d_t \mu_{j,k} + \mu_{j,k} d_t\lambda^{\Phi}_{j,k}  \right) \end{equation}

In the constant transport coefficient approximation, Eq. \ref{dege6} reads:

\begin{equation}\label{dege7} d_tG = -\Psi d_tT + \sum_i {\mu_i d_tC_i} + \sum_j \sum_k (\lambda^{\Phi}_{j,k})^{2}  \lambda^{X}_j \mu_{j,k} \, d_t\mu_{j,k}\end{equation}

Defining $L_{j,k} = \frac{(\lambda^{\Phi}_{j,k})^{2} \lambda^{X}_j}{2}$

\begin{equation}\label{dege8} d_tG = -\Psi d_tT + \sum_i {\mu_i d_tC_i} + \sum_j \sum_k  L_{j,k} \, d_t \mu_{j,k}^{2}\end{equation}

It is possible to see from Eq. \ref{dege8} that genetic transcription could be characterized as a \emph{second-order} effect, this raises from the fact that the actual mechanism of gene expression is regulation by other gene products such as enzymes and transcription factors.\\

\begin{equation}\label{dege9} d_tG = -\Psi \, d_tT + \sum_i {\mu_i d_tC_i} + \sum_j \sum_k  L_{j,k} \, d_t \mu_{j,k}^{2} \end{equation}

As we have stated, fluorescence intensity signals as measured by, for example, Microarray experiments (i.e. gene chips) are the usual technique to acquire information about the concentration of a given gene under certain cellular conditions. From Eq. \ref{specint2}, the concentration of a given gene-probe (with hybridization energy $\Delta G_i$) is a function of the intensity as follows:

\begin{equation}\label{intense} c_i = \frac{\varphi_i}{A_i  e^{- \beta \Delta G_i} - \varphi_i e^{- \beta \Delta G_i} }\end{equation}

It is not unreasonable to consider that the local single-species \emph{energy of formation} for a given mRNA transcript, (i.e. the partial chemical potential $\mu_i$ in Eq. \ref{dege9}) has the same (absolute) value as the chemical potential of hybridization for the same mRNA species, as given by Eq. \ref{muifinal} such that $\mu_i = + RT /c_i$ could be used in the thermodynamical characterization of gene expression as given by Eq. \ref{dege9}. If we insert Eq. \ref{intense} into Eq. \ref{muifinal} we get:

\begin{equation}\label{intenmu} \mu_i = \frac{RT \left( A_i e^{- \beta \Delta G_i} - \varphi_i e^{- \beta \Delta G_i}\right)}{\varphi_i}\end{equation}

Now, by taking the time derivative of Eq. \ref{intense}:

\begin{equation}\label{intenci} \frac{d c_i}{dt} = \frac{A_i \, e^{- \beta \Delta G_i}}{\left( A_i \, e^{- \beta \Delta G_i} - \varphi_i \, e^{- \beta \Delta G_i}\right)^{2}} \left[\frac{d \varphi_i}{dt}\right] \end{equation}

By substitution of Eq. \ref{intenmu} and Eq. \ref{intenci} into Eq. \ref{dege9}:

\begin{equation}\label{dege10} d_tG = -\Psi \, d_tT + \sum_i \frac{A_i \, e^{- \beta \Delta G_i}}{\beta \varphi_i \left( A_i \, e^{- \beta \Delta G_i} - \varphi_i \, e^{- \beta \Delta G_i}\right)} \, d_t \varphi_i + \sum_j \sum_k  L_{j,k} \, d_t \mu_{j,k}^{2} \end{equation}

If we define $\Gamma_i =  \frac{A_i \, e^{- \beta \Delta G_i}}{\beta \varphi_i \left( A_i \, e^{- \beta \Delta G_i} - \varphi_i \, e^{- \beta \Delta G_i}\right)}$ as the \emph{thermodynamic} conjugate variable to the probe intensity $\varphi_i$ we obtain:\\

\begin{equation}\label{dege11} d_tG = -\Psi \, d_tT + \sum_i \Gamma_i \, d_t \varphi_i + \sum_j \sum_k  L_{j,k} \, d_t \mu_{j,k}^{2} \end{equation}

\section{Results and Discussion}
Let us examine in some detail the structure of Eq. \ref{dege11}. In the isothermic, non-regulated steady state (i.e. $d_t G = 0$, $d_tT =0$, $d_t \mu_{j,k}^{2} = 0 \; \forall j,k$), Eq. \ref{dege11} is nothing but a formal non-equilibrium extension of the Gibbs-Duhem relation $\sum_i \Gamma_i \, d_t \varphi_i = 0$. Without any gene regulatory mechanism, and without explicit dissipation, the energetics of gene expression within a cell are just the ones of a non-interacting dilute mixture of its components (in this case the different mRNA transcripts). A more realistic case is the regulated, isothermal steady state given by: $d_t G = 0$, $d_tT =0$ and at least some $d_t \mu_{j,k}^{2} \neq 0$. This is the more interesting case that one can compare with actual gene transcription experiments nowadays. This is so because, on one hand, due to the specific nature of nucleic acids (both DNA and RNA suffer thermal decay) and also due to physiological conditions; temperature changes are subtle or negligible within the living cell or inside a realistic biological assay. \\

The steady state condition is more of a \emph{present-time situation} than a definitive limitation. Most dynamic gene expression studies nowadays are studied as time series (or \emph{time-courses} in the biomedical language) with time-steps dictated by economical or pharmacological and not by biophysical reasons. Typically, the smaller time-steps are of the order of minutes if not hours or days. Regulatory changes can be thus measured just in their steady-state mean-field contributions (coarse grained in space and time) and not in their whole dynamical complexity. Of course, as the costs of Microarray processing lower and as the technologies advance, one expects to see better resolution time series for transcriptional dynamics.\\

Let us then consider the regulated isothermal steady-state version of Eq. \ref{dege11}, namely:

\begin{equation}\label{dege12} \sum_i \Gamma_i \, d_t \varphi_i + \sum_j \sum_k  L_{j,k} \, d_t \mu_{j,k}^{2} = 0 \end{equation}

One could see that changes in the mRNA concentration of gene $i$ as measured by its probe intensity $\varphi_i$ could depend \emph{not only in their own} characteristic thermodynamical parameters ($A_i$, $\Delta G_i$, and $T$) but also on other mRNA transcript (say $n$) via a coupling given by a term $L_{n,i} \, \mu_{n,i}^{2}$. In that case one says that the $n$-th gene regulates the $i$-th gene, or that $n$ is a \emph{transcription factor} for $i$ (conversely $i$ is a \emph{transcriptional target} of $n$). \\

In order to give a concrete example (for the sake of clarity), we will consider the irreversible thermodynamic coupling that sets the process of transcriptional regulation between two genes $Genes = \{1,2\}$. In this case we will assume that gene number $1$ is a transcription factor for gene number $2$ and that gene $1$ is non-regulated (i.e. gene $1$ is not a target for any TF). This means that $\mu_{1,2} \neq 0$ and that $\mu_{1,1} = \mu_{2,1} =\mu_{2,2} = 0$. In this case Eq. \ref{dege12} will read:

\begin{equation}\label{gammafinal} \Gamma_1 \, d_t \varphi_1 + \Gamma_2 \, d_t \varphi_2 + L_{1,2} \, d_t \mu_{1,2}^{2} = 0 \end{equation}

To make explicit calculations from experimental data we will consider \textit{SYK}, the transcript responsible for the synthesis of \emph{spleen tyrosine kinase} as gene 1 and \textit{IL2RB} or \emph{interleukin 2 receptor, beta} as gene 2. \textit{SYK} is well known for being a strong inducer of gene transcription, specially in the case of the beta domain interleukin 2 receptor \cite{syk}. Also, there is a strong evidence indicating the possible role of these two genes in the course of the so-called C-MYC network of reactions, a very important, cancer-related biochemical pathway.\\

The values of the parameters could be calculated as follows. According with the algorithm developed by Lu, et al \cite{lu} and described by Carlon, et al (cf. Table 1 of reference \cite{carlon}) it is possible to obtain suitable values for $\Delta G_1 = 483.55$ kcal/mol and $\Delta G_2 = 463.05$ kcal/mol (see Table 1). From these values, we can calculate $A_1$ and $A_2$ from Eq. \ref{specint2} following saturation measurements in the \emph{latin square} experiments \cite{affy,lu,carlon}. In this case $A_1 = 5513$ intensity units/mol and $A_2 = 1105$ intensity units/mol (see figures 1 and 2). \\

Given these parameters, from a time-course GEA it is possible to calculate both $\Gamma_1= \Gamma_1(\varphi_1)$ and $\Gamma_2= \Gamma_2(\varphi_2)$, and via $\varphi_1(t)$ and $\varphi_2(t)$ we could as well obtain the time evolution for $\mu_{1,2}$, hence characterizing in a complete form the transcriptional regulation for this simple (almost trivial from the biological standpoint) \emph{gene switch}.\\

Taking the aforementioned values, we have the following expressions for the thermodynamic functions in terms of the experimentally measurable intensities (in all cases a physiological temperature of $T = 37 ^{o} \,C$ is assumed), hence $\beta = 1.622507 \times 10 ^{-6}$ mol kcal$^{-1}$,  $e^{-\beta \Delta G_1} = 0.99922$, $A_1 \times e^{-\beta \Delta G_1} = 5508.67950$ intensity units/mol;  also  $e^{-\beta \Delta G_2} = 0.99925$, and $A_2 \times e^{-\beta \Delta G_2} = 1104.17014$ intensity units/mol.\\

Calculating the intensity-dependent chemical potentials we obtain, from Eq. \ref{intenmu}  kcal/mol:

\begin{equation}\label{muin1} \mu_1 = \frac{3.395 \times 10 ^{9} - 6.158 \times 10 ^{5} \; \varphi_1}{\varphi_1} \end{equation}

and

\begin{equation}\label{muin2} \mu_2 = \frac{6.805 \times 10 ^{8} - 6.159 \times 10 ^{5}\; \varphi_2}{\varphi_2} \end{equation}

As we could seed from Eq. \ref{muin1} and \ref{muin2} (Figure 3), there is a difference in the transcriptional behavior of gene 1 (\textit{SYK}) which is a transcription factor and gene 2 (\textit{IL2RB}) which is not (and, in fact is a transcriptional target). The maximum intensity (related to a maximum concentration peak) attainable in both cases in the spontaneous regime is of 5513 intensity units for \textit{SYK}, whereas in the case of \textit{IL2RB} is of just 1105 intensity units. This means that, in order for \textit{IL2RB} to be produced at higher rates, the presence of chemical environment modifications (e.g via transcription factors) is needed.\\

In a very straightforward way (similar to our $\mu_i$ calculations) we are now able to calculate expressions for $\Gamma_1$ and $\Gamma_2$ as follows (see Figure 4).

\begin{equation}\label{Gamma1} \Gamma_1 = \frac{5508.6795}{0.008938 \; \varphi_1 - 1.6098 \times 10 ^{-6} \; \varphi_1^{2}}\end{equation}

\begin{equation}\label{Gamma2} \Gamma_2 = \frac{1104.17014}{0.001795 \; \varphi_2 -  1.6243 \times 10 ^{-6}\; \varphi_2^{2}}\end{equation}

If we substitute Eq. \ref{Gamma1} and \ref{Gamma2} into Eq. \ref{gammafinal} we obtain:

\begin{eqnarray} \label{gammafinal2}
  - L_{1,2}\,d_t \mu_{1,2}^{2} &=& \frac{5508.6795}{0.008938 \; \varphi_1 - 1.6098 \times 10 ^{-6} \; \varphi_1^{2}} \; d_t \varphi_1 \nonumber \\
  &+& \frac{1104.17014}{0.001795 \; \varphi_2 -  1.6243 \times 10 ^{-6}\; \varphi_2^{2}} \; d_t \varphi_2
\end{eqnarray}

Integrating

\begin{eqnarray}\label{gammafinal3}
- L_{1,2}\mu_{1,2}^{2} &=&  616321.2687 \ln \left| \frac{-1.6098\times 10 ^{-6} \; \varphi_1}{0.008938 -1.6098\times 10 ^{-6} \;\varphi_1} \right| \nonumber \\
  &+& 615, 136.5683 \ln \left| \frac{-  1.6243 \times 10 ^{-6}  \;\varphi_2}{0.001795 - 1.6243 \times 10 ^{-6} \; \varphi_2}\right|
\end{eqnarray}

Taking experimental values of $\varphi_1(t)$ and $\varphi_2(t)$, Eq. \ref{gammafinal3} could be solved for $\mu_{1,2}(t)$.  As we already stated, both \textit{SYK} and \textit{IL2RB} are involved in the transcriptional network related to the C-MYC pathway which is very important in the development of cancer. \\

In order to capture more subtle regulatory dynamics one will need experiments with a large number of smaller time-step measurements, but in principle one is able to observe detailed patterns even within this very simple thermodynamic model.\\

Interestingly, for this single gene switch it is also possible to calculate the dependency of the transcriptional regulation coupling $\mu_{1,2}$ on the particular cellular environment by solving Eq. \ref{gammafinal3} for the same two genes under different phenotypical conditions (e.g. cancer versus normal cells, treated vs untreated diseased cells, etc.). The systematic study of such thermodynamic cellular-context transcription regulation theory seems to be a promising research area in the non-equilibrium thermodynamics of biosystems. In conclusion, we have showed here that a non-equilibrium thermodynamical description of cell-level transcriptional regulation could be formulated in terms of experimentally measurable quantities, and that essential features of gene regulatory dynamics could be studied with it. The model has been progressively simplified to match with todays technological and practical limitations, nevertheless these simplifications are not necessary in principle, and can be eliminated when better experimental resolution (specially with regards to more samples and time-points) could be attained.

\pagebreak

{\small \textbf{Table 1}: \textit{Thermodynamic data for gene transcripts included in the Latin Square experiments. $\Delta G_{tr}$ at $37^{o}C$ are calculated according to reference \cite{carlon}}}.\\

{\tiny
\begin{tabular}{ccccc}
\hline \\
probeset\_key	&	Gene	&	Gene Name	&	Transcription Factor activity	&	$\Delta G_{tr}$ at $37^{o}C$ (Kcal/mol)	\\
\\
\hline \\
203508\_at	&	TNFRSF1B	&	tumor necrosis factor receptor superfamily, member 1B	&		&	502.52	\\
204563\_at	&	SELL	&	selectin L	&		&	446.34	\\
204513\_s\_at	&	ELMO1	&	engulfment and cell motility 1	&		&	471.35	\\
204205\_at	&	APOBEC3G	&	apolipoprotein B mRNA editing enzyme, catalytic polypeptide-like 3G	&	Reverse TF	&	 477.38	\\
204959\_at	&	MNDA	&	myeloid cell nuclear differentiation antigen	&	TF Regulation	&	433.97	\\
207655\_s\_at	&	BLNK	&	B-cell linker	&		&	436.55	\\
204836\_at	&	GLDC	&	glycine dehydrogenase (decarboxylating)	&		&	468.99	\\
205291\_at	&	IL2RB	&	interleukin 2 receptor, beta	&		&	463.05	\\
209795\_at	&	CD69	&	CD69 molecule	&		&	398.72	\\
207777\_s\_at	&	SP140	&	SP140 nuclear body protein	&	TF activity	&	700.57	\\
204912\_at	&	IL10RA	&	interleukin 10 receptor, alpha	&		&	474.56	\\
205569\_at	&	LAMP3	&	lysosomal-associated membrane protein 3	&		&	636.01	\\
207160\_at	&	IL12A	&	interleukin 12A	&		&	453.31	\\
205692\_s\_at	&	CD38	&	CD38 molecule	&		&	569.56	\\
212827\_at	&	IGHM	&	immunoglobulin heavy constant mu	&		&	482.07	\\
209606\_at	&	PSCDBP	&	    cytohesin 1 interacting protein	&		&	458.19	\\
205267\_at	&	POU2AF1	&	POU class 2 associating factor 1	&	TF Regulation	&	473.5	\\
204417\_at	&	GALC	&	galactosylceramidase	&		&	410.95	\\
205398\_s\_at	&	SMAD3	&	SMAD family member 3	&	TF activity + Binding	&	465.08	\\
209734\_at	&	NCKAP1L	&	NCK-associated protein 1-like	&		&	716.67	\\
209354\_at	&	TNFRSF14	&	tumor necrosis factor receptor superfamily, member 14	&		&	782.09	\\
206060\_s\_at	&	PTPN22	&	protein tyrosine phosphatase, non-receptor type 22	&		&	414.77	\\
205790\_at	&	SKAP1	&	src kinase associated phosphoprotein 1	&	TF	&	452.57	\\
200665\_s\_at	&	SPARC	&	secreted protein, acidic, cysteine-rich (osteonectin)	&		&	443.24	\\
207641\_at	&	TNFRSF13B	&	tumor necrosis factor receptor superfamily, member 13B	&	TF inducer	&	481.8	\\
207540\_s\_at	&	SYK	&	spleen tyrosine kinase	&	TF inducer	&	483.55	\\
204430\_s\_at	&	SLC2A5	&	solute carrier family 2 (facilitated glucose/fructose transporter), member 5	&		 &	488.56	\\
203471\_s\_at	&	PLEK	&	pleckstrin	&		&	459.23	\\
204951\_at	&	RHOH	&	ras homolog gene family, member H	&	TF Regulation	&	467.94	\\
207968\_s\_at	&	MEF2C	&	myocyte enhancer factor 2C	&	TFact, RNAPol ind	&	472.81	\\
\\
\hline \\
\end{tabular}}

\begin{figure}[h]
\centerline{\psfig{file=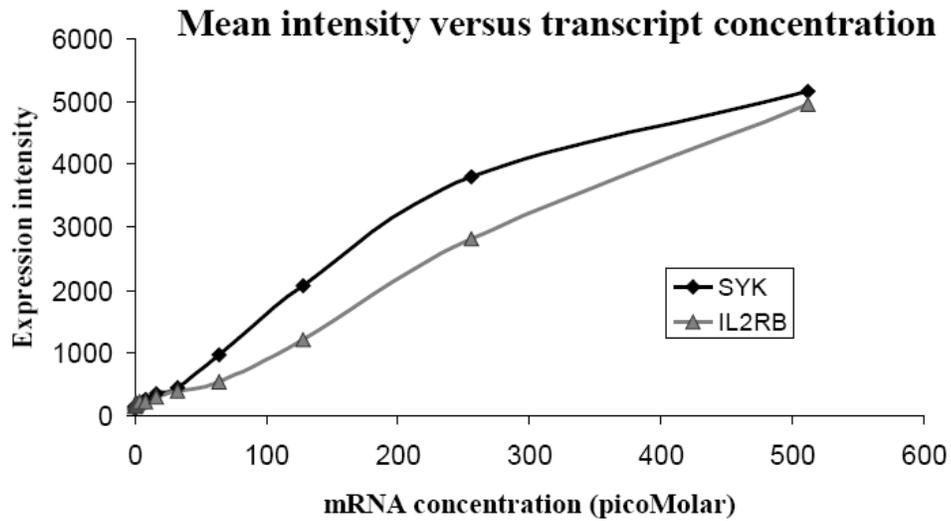,width=14cm,angle=0}}
\caption{Gene expression intensity as a function of mRNA concentration for \textit{SYK} and \textit{IL2RB}}
\label{Figure 1}
\end{figure}

\begin{figure}[h]
\centerline{\psfig{file=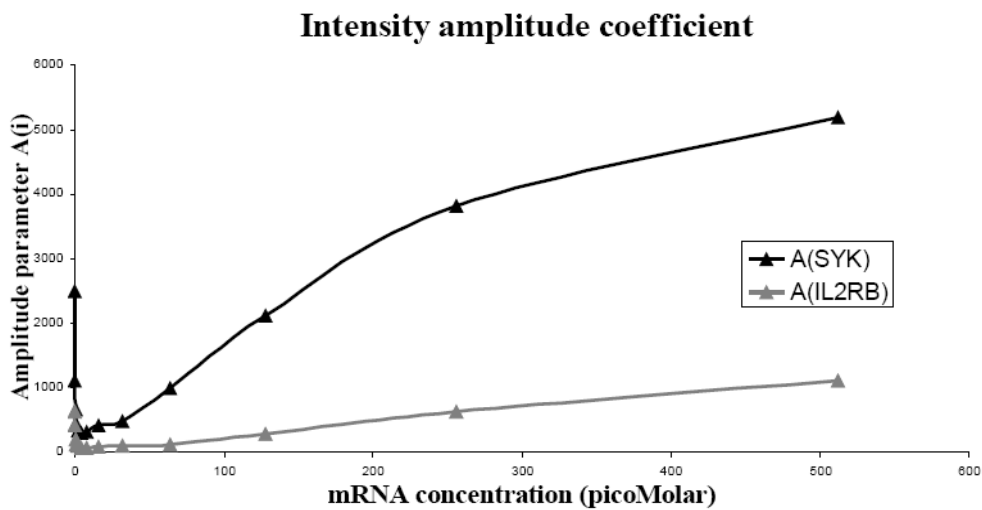,width=14cm,angle=0}}
\caption{Intensity amplitude coefficient as a function of mRNA concentration for \textit{SYK} and \textit{IL2RB}}
\label{Figure 2}
\end{figure}

\begin{figure}[h]
\centerline{\psfig{file=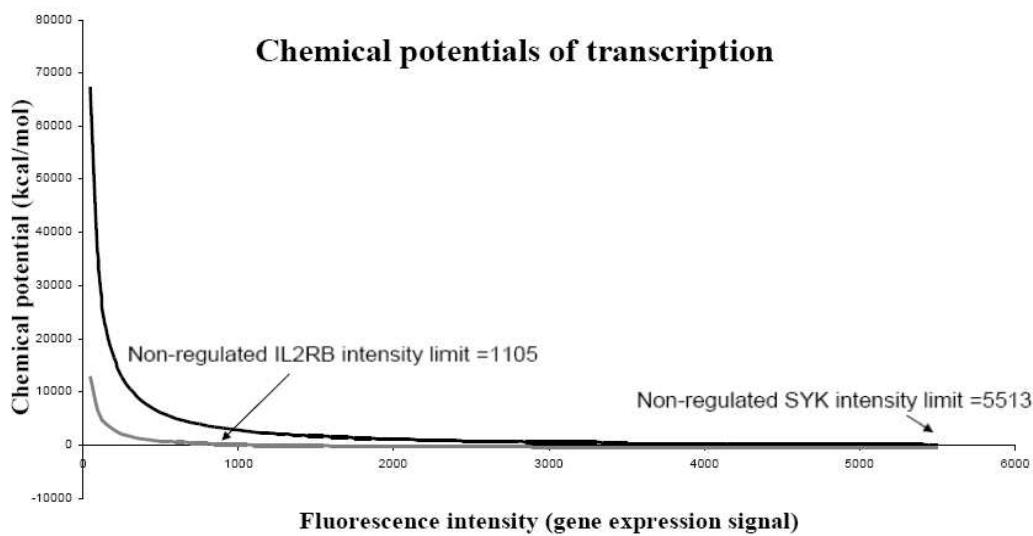,width=14cm,angle=0}}
\caption{Individual chemical potentials for non-regulated transcription $\mu_{SYK}$ and $\mu_{IL2RB}$ as a function of gene expression intensity}
\label{Figure 3}
\end{figure}

\begin{figure}[h]
\centerline{\psfig{file=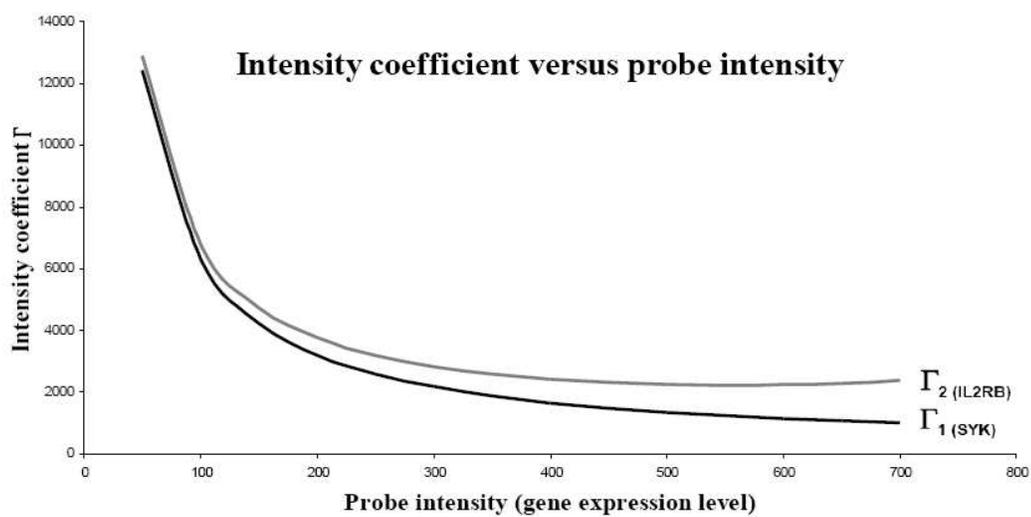,width=14cm,angle=0}}
\caption{Intensity parameters for non-regulated transcription $\Gamma_{SYK}$ and $\Gamma_{SYK}$ as a function of gene expression intensity}
\label{Figure 4}
\end{figure}

\end{document}